\documentclass[acmsmall]{acmart}
\usepackage{enumitem}
\usepackage{listings}
\usepackage[draft=true]{minted}
\usepackage{tikz}
\usetikzlibrary{positioning}
\usetikzlibrary{calc}
\usetikzlibrary{arrows.meta}
\usepackage{caption}
\usepackage{subcaption}
\usepackage{wrapfig}
\usepackage{microtype}
\AtBeginDocument{%
  \providecommand\BibTeX{{%
    \normalfont B\kern-0.5em{\scshape i\kern-0.25em b}\kern-0.8em\TeX}}}

\newcommand{\alice}{Alice}
\newcommand{\bob}{Bob}
\setcopyright{acmcopyright}
\copyrightyear{2022}
\acmYear{2022}
\acmConference[ProLaLa '23]{Workshop on Programming Languages and the}{January 15, 2023}{Boston, MA}

\begin{document}






\title[Exploring Consequences of Privacy Policies]{Exploring Consequences of Privacy Policies \\ with Narrative Generation via Answer Set Programming}

\author{Chinmaya Dabral}
\email{csdabral@ncsu.edu}
\orcid{}
\affiliation{
  \institution{North Carolina State University}
  \streetaddress{}
  \city{Raleigh}
  \state{NC}
  \country{USA}
  \postcode{}
}
\author{Emma Tosch}
\email{e.tosch@northeastern.edu}
\orcid{0000-0002-2333-8034}
\affiliation{
  \institution{Northeastern University}
  \streetaddress{}
  \city{Boston}
  \state{MA}
  \country{USA}
  \postcode{}
}
\author{Chris Martens}
\email{c.martens@northeastern.edu}
\orcid{}
\affiliation{
  \institution{Northeastern University}
  \streetaddress{}
  \city{Boston}
  \state{MA}
  \country{USA}
  \postcode{}
}

\renewcommand{\shortauthors}{Dabral et al.}

\begin{abstract}
Informed consent has become increasingly salient for data privacy and its regulation. Entities from governments to for-profit companies have addressed concerns about data privacy with policies that enumerate the conditions for personal data storage and transfer. However, increased enumeration of and transparency in data privacy policies has not improved end-users' comprehension of how their data might be used: not only are privacy policies written in legal language that users may struggle to understand, but elements of these policies may compose in such a way that the consequences of the policy are not immediately apparent. 

We present a framework that uses Answer Set Programming (ASP) --- a type of logic programming --- to formalize privacy policies. Privacy policies thus become constraints on a narrative planning space, allowing end-users to forward-simulate possible consequences of the policy in terms of \emph{actors} having \emph{roles} and taking \emph{actions} in a \emph{domain}.
We demonstrate through the example of the Health Insurance Portability and Accountability Act (HIPAA) how to use the system in various ways, including asking questions about possibilities and identifying which clauses of the law are broken by a given sequence of events.
\end{abstract}

\maketitle

\section{Introduction}
The rules, laws, and norms for releasing private information are constantly evolving; consequently,
individuals are asked to agree to increasingly complex laws and terms of service that they may not understand. 
Despite the stakes, research shows that even well-educated users have trouble understanding privacy policies \cite{proctor2008examining}, creating an \textit{information asymmetry} between providers of these services and the user, where the user is not fully aware of the consequences of their consent,  leading them to agree to release information under circumstances they could not have anticipated~\cite{acquisti2007can}. 

Prior formalization efforts have focused on verifying the compliance of data-sharing software with the law. Unfortunately formalization has not focused on end-user understanding of the complex pathways by which their data may be shared. We present a novel approach to formalization of privacy policies that can be used to \emph{explain} the consequences of privacy policies via narrative. This work focuses on the technical backend of such a system and proposes a possible rendering. 

\paragraph{Motivating Example: HIPAA via Narrative Constraints}
The Health Insurance Portability and Accountability Act (HIPAA) is a US law enacted in 1996. It was created to modernize and regulate the flow of healthcare information, and protect patient privacy. 
HIPAA is well-studied in the privacy compliance literature, has a history of interpretation, and is an exemplar of modular and non-trivial privacy legislation. HIPAA has previously been formalized via Prolog in service of the verifying software-based data sharing and we leverage this prior work in our encoding~\cite{lam2009formalization, lam2012declarative}. Here we focus on the part of HIPAA that regulates the transmission of protected health information (PHI) by healthcare entities and use our system to answer the following questions:

\begin{enumerate}[label={\textbf{Q\arabic*}}, leftmargin=*]
    \item\label{q:alice-phi-third-parties} \alice{} asks: can hospitals sell my PHI to third parties?
    \item\label{q:alice-phi-third-parties-no-authorization} \alice{} asks: when can hospitals sell my PHI to third parties \emph{without my authorization}?
    \item\label{q:bob-employer-phi} \bob{} asks: when can my employer obtain my PHI?
\end{enumerate}
\section{Background and System Design}
\label{sec:background}

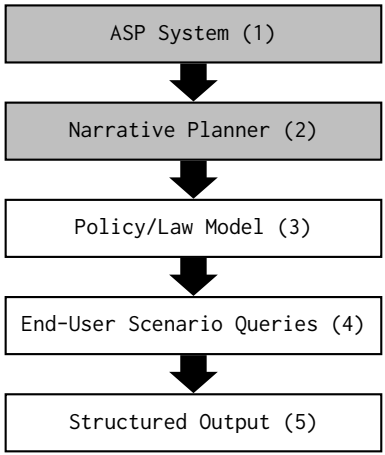
\begin{wrapfigure}[28]{L}{0.4\textwidth}
\vspace{-1em}
\centering
\begin{tikzpicture}
\tikzstyle{every node}=[rectangle, draw, thick, font=\small\ttfamily, minimum width=5cm, minimum height=0.75cm]
    \node[fill=lightgray] (asp) at (0,0) {ASP System (1)};
    \node[fill=lightgray, below=5mm of asp] (narrative) {Narrative Planner (2)};
    \node[below=5mm of narrative] (law) {Policy/Law Model (3)};
    \node[below=5mm of law] (query) {End-User Scenario Queries (4)};
    \node[below=5mm of query] (output) {Structured Output (5)};
    \draw[ -{Triangle[width = 18pt, length = 8pt]}, line width = 10pt] (asp.south) -- (narrative.north);
    \draw[ -{Triangle[width = 18pt, length = 8pt]}, line width = 10pt] (narrative.south) -- (law.north);
    \draw[ -{Triangle[width = 18pt, length = 8pt]}, line width = 10pt] (law.south) -- (query.north);
    \draw[ -{Triangle[width = 18pt, length = 8pt]}, line width = 10pt] (query.south) -- (output.north);
\end{tikzpicture}
\caption{\label{fig:system}The end-to-end architecture of our privacy exploration system. Greyed out blocks represent prior work: the Clingo ASP programming system~\cite{gebser2008user} provides the general language and solver, but does not include the primitives necessary to reason about narrative. We leverage the general-purpose narrative planner of \cite{dabral2020generating} to encode rules and constraints pertaining to time and actors. This work illustrates one possible instantiation of a privacy law: HIPAA, as well as several end-user queries and structured graphical output. }
\end{wrapfigure}
We formalize privacy policies using Clingo, an Answer Set Programming (ASP) system \cite{gebser2008user}. ASP is a declarative programming technique where logic programs are specified in terms of generative rules, facts, and constraints in a syntax similar to Prolog. A constraint solver then attempts to find stable models (answer sets) for the logic program \cite{gelfond1988stable}. 
ASP differs from classical logic programming via its ``choice rules,'' which allow multiple possible worlds to be consistent with a given program; syntactically, the head of choice rules in ASP may be an arbitrary-sized set of ground atoms. 

\emph{Narrative generation} is the creation of causally-related and temporally ordered sequences of events that form a \emph{story} having certain properties; it is commonly framed as an AI planning task. Narratives depicting plausible real-world scenarios that are consistent with privacy policies would serve well as examples to educate users, while also addressing the context-dependence of user perceptions of privacy \cite{nissenbaum2004privacy}. 

A \emph{narrative planner} is software that generates a partial order of events. Implementing a planner in ASP opens up the possibility of using the rich constraints provided by the system for sculpting the possibility space of generated narratives.
Our narrative planner is general purpose and supports \emph{intentional actors} (i.e., agents that act with purpose) and \emph{conflicts} (i.e., heterogenous intents that may not be satisfiable). For now we focus strictly on intentional and cooperative actors.

\section{Encoding Privacy Policies}
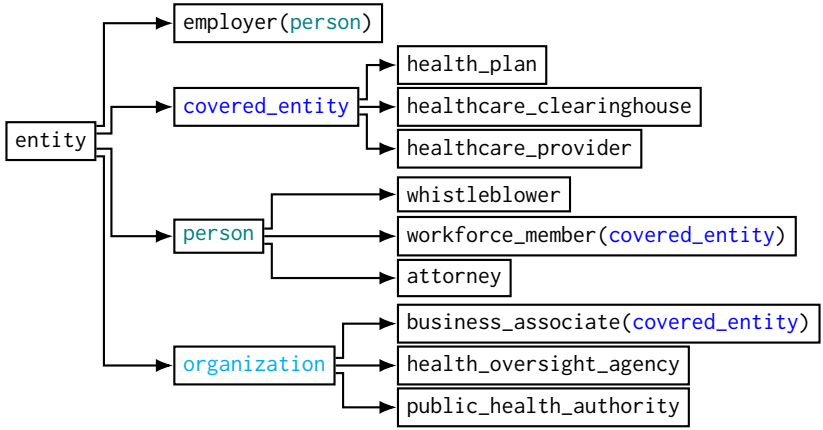
\begin{figure}
\centering
\begin{tikzpicture}
\tikzstyle{every node}=[rectangle, draw, thick, font=\small\ttfamily]
\node[anchor =west] (pha) {public\_health\_authority};
\node[above = 5.5mm of pha.west, anchor=west] (hoa) {health\_oversight\_agency};
\node[above = 5.5mm of hoa.west, anchor=west] (ba) {business\_associate(\textcolor{blue}{covered\_entity})};
\node[above = 6mm of ba.west, anchor=west] (a) {attorney};
\node[above = 5.5mm of a.west, anchor=west] (wm) {workforce\_member(\textcolor{blue}{covered\_entity})};
\node[above =5.5mm  of wm.west, anchor=west] (w) {whistleblower};
\node[above = 6mm of w.west, anchor=west] (hp) {healthcare\_provider};
\node[above =5.5mm of hp.west, anchor=west] (hc) {healthcare\_clearinghouse};
\node[above =5.5mm of hc.west, anchor=west] (hpl) {health\_plan};
\node[left=0.5cm of hc.west] (ce) {\textcolor{blue}{covered\_entity}};
\node[above=1.1cm of ce.west, anchor=west] (employer) {employer(\textcolor{teal}{person})};
\node[anchor=west] at (wm.west -| ce.west) (person) {\textcolor{teal}{person}};
\node[anchor=west] at (hoa.west -| ce.west) (org) {\textcolor{cyan}{organization}};
\node[left=of person] at (-3,3.5) (entity) {entity};
\draw[thick, -Latex] ($(org.east) + (0, -1mm)$) -- ($(org.east) + (1mm, -1mm)$) --  ($(org.east) + (1mm, -1mm) - (pha.west)$) |- (pha.west);
\draw[thick, -Latex] (org.east) -- (hoa.west);
\draw[thick, -Latex] ($(org.east) + (0,  1mm)$) -- ($(org.east) + (1mm, 1mm)$)  |- (ba.west);
\draw[thick, -Latex] (person.east) -- (wm.west);
\draw[thick, -Latex] ($(person.east) + (0, -1mm)$)  -- ($(person.east) + (1mm, -1mm)$) |- (a.west);
\draw[thick, -Latex] ($(person.east) + (0,  1mm)$) -- ($(person.east) + (1mm, 1mm)$)  |- (w.west);
\draw[thick, -Latex] (ce.east)--(hc.west);
\draw[thick, -Latex] ($(ce.east) - (0, 1mm)$) -- ($(ce.east) + (1mm, -1mm)$) |- (hp.west);
\draw[thick, -Latex] ($(ce.east) + (0, 1mm)$) -- ($(ce.east) + (1mm, 1mm)$) |- (hpl.west);
\draw[thick, -Latex] ($(entity.east) + (0, 2mm)$) -- ($(entity.east) + (1mm, 2mm)$) |- (employer.west);
\draw[thick, -Latex] ($(entity.east) + (0, 1mm)$) -- ($(entity.east) + (2mm, 1mm)$) |- (ce.west);
\draw[thick, -Latex] ($(entity.east) + (0, -1mm)$) -- ($(entity.east) + (2mm, -1mm)$) |- (person.west);
\draw[thick, -Latex] ($(entity.east) + (0, -2mm)$) -- ($(entity.east) + (1mm, -2mm)$) |- (org.west);
\end{tikzpicture}
\caption{Partial role hierarchy. Note that certain roles are parameterized. These parameterizations represent constraints on the role. For example, ``participant" from the quoted clause of \S164.506(c)(5) refers to a \texttt{\textcolor{blue}{covered\_entity}} that participates in a specific \texttt{\textcolor{cyan}{organization}}.}
\label{fig:role_hierarchy}
\end{figure}

The first step in our framework is to encode the relevant \emph{roles} that an actor may assume and the \emph{actions} an actor inhabiting a role may take. This corresponds to implementing Box 3 of Figure~\ref{fig:system}. In our example context, actions that correspond the transmission of patient data as documented in HIPAA clauses become narrative constraints. 

\paragraph{Encoding a Role Hierarchy}

Most clauses in HIPAA are conditioned on the role of the involved entities. For instance, \S164.506(c)(5) mentioned four roles, emboldened for emphasis:
\begin{quote}
A \textbf{\textcolor{blue}{covered entity}} that participates in an \textbf{\textcolor{cyan}{organized health care arrangement}} may disclose protected health information about an \textbf{individual} to other \textbf{participants} in the organized health care arrangement for any health care operations activities of the organized health care arrangement.    
\end{quote}

The roles are relational and form a hierarchy --- e.g., ``\textcolor{blue}{covered entity}'' is further defined in \S160.103 as either a health care provider (doctors, clinics, etc.), a health plan, or a health care clearinghouse. 
Care must be taken to ensure that the hierarchy expresses strict ``is-a'' relationships; the \texttt{\textcolor{blue}{covered\_entity}} role is an example of this enforcement. Note that a \texttt{\textcolor{blue}{covered\_entity}} can be a \texttt{\textcolor{teal}{person}} (e.g., a doctor) or an \texttt{\textcolor{cyan}{organization}} (e.g., a hospital). 
For example, Seattle Grace Hospital is a \texttt{\textcolor{blue}{covered\_entity}} by virtue of \texttt{healthcare\_provider(sgh)}. 
Then Dr. Cristina Yang holds at least two roles: she is a \texttt{\textcolor{blue}{covered entity}} by virtue of \texttt{healthcare\_provider(yang)} and a \texttt{\textcolor{teal}{person}} by virtue of \texttt{workforce\_member(sgh)(yang)}.\footnote{For simplicity we are treating all predicates as curried; this is not Clingo syntax.} Figure~\ref{fig:role_hierarchy} depicts a a partial graph of the role hierarchy. 

\paragraph{Encoding Permissible Actions}
Actions in the generated narrative are carried out by \textit{agents} with roles. An agent with a given role is also automatically assigned all parent roles in the hierarchy. Additionally, an agent may also take on an unconnected role. For instance, a doctor may also take the role of patient, with the obvious restriction that an organization cannot also be a person. These roles may be assigned manually as part of the initial conditions, or automatically by the planner in order to satisfy a given narrative constraint.

HIPAA defines many actions that can be performed by agents. For instance, \S164.512(f)(3) states:
\begin{quote}
Except for disclosures
required by law as permitted by
paragraph (f)(1) of this section, a
covered entity may \textbf{\textcolor{red}{disclose}} protected
health information in response to a
law enforcement official's \textbf{request} for
such information about an individual
who is or is \textbf{suspected} to be a victim
of a crime, other than disclosures that
are subject to paragraph (b) or (c) of
this section, if: (i) The individual \textbf{\textcolor{orange}{agrees}} to the
disclosure; or (ii) The covered entity is unable to
\textbf{\textcolor{purple}{obtain}} the individual's agreement
because of incapacity or other
emergency circumstance...
\end{quote}

There are some \emph{latent} actions that an expert must encoded in order to form a coherent narrative. For instance, to \textcolor{red}{disclose} protected health information, the \textcolor{blue}{covered entity} must first \textit{\textcolor{magenta}{acquire}} it, and the acquired information must be about the specific patient in question. There are many ways to \textcolor{magenta}{acquire} information: e.g., via doctor's \textit{visit}, or any of a number of possible real world actions. We need to encode at least one such action in our model.
Similarly, the ``incapacity" must also be the consequence of an event, i.e., we must be able to negate an action.

Fluents track the state of the world and  causally link actions. In \S164.512(f)(3), \textcolor{orange}{agreeing} to the disclosure must occur \emph{after} the attempt to \textcolor{purple}{obtain} the individual's agreement is made. We encode this attempt as a fluent. 
Roles can also be tracked as fluents, allowing them to change through the narrative. 
Figure~\ref{fig:listing-action-encoding} gives an example action encoding.

\begin{figure}\centering
\begin{minted}{prolog}
% Giving treatment to a person is one way to 
% acquire their health info
possible(I, A, al(Person, al(CE))):-
    id(I),
    A=action(get_treatment, Person, CE),
    role(Person, person),
    role(CE, covered_entity).

initiates(I, A, has_info(CE,Person,phi,I)):-
    A=action(get_treatment, Person, CE),
    happens(I, A, _).
\end{minted}
\caption{\label{fig:listing-action-encoding}Example action encoding for one way to \textcolor{magenta}{acquire} information. The predicates \texttt{possible} and \texttt{intiates} are part of the narrative generator. Each encoded privacy policy must define roles and actions; \texttt{get\_treatment} is specific to HIPAA. }
\end{figure}
\begin{figure}[t]
\includegraphics[width=0.8\textwidth]{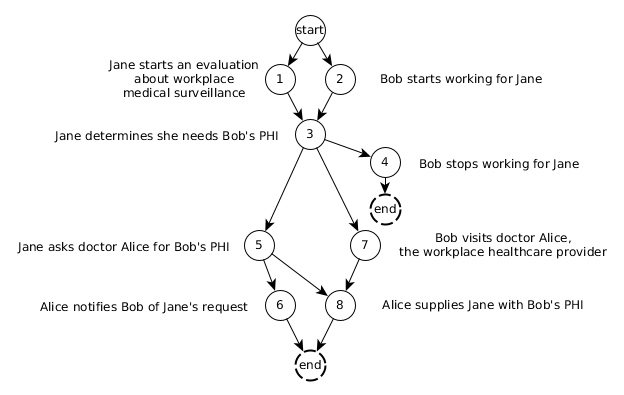}
\caption{Templated explanation of narrative output for ~\ref{q:bob-employer-phi}. Paths through the graph represent a partial ordering of possible events that could lead to Bob's employer obtaining his PHI.}
\label{fig:antagonistic}
\end{figure}

\paragraph{Encoding Constraints and Additional Objects}
In addition to actors which have roles and can take actions, our system requires reasoning about abstract objects, such as PHI. For instance, a clause applicable to ``private health information" must also apply to ``psychotherapy notes" (\S164.508(a)(2)), since latter is a type of former.
Thus we encode other objects in our system using a hierarchy similar to the role heirarchy.

Clauses allow or disallow transmission of protected information based on certain conditions. Additionally, they may impose future obligations on entities. Here we borrow the encoding scheme of \citet{lam2009formalization}, with an additional term to allow future obligations to be checked ($obligation_R$); this rule is necessary to ensure the coherence of narratives. Table~\ref{tab:transmission-rules} lists the full set of transmission rules.
\begin{table}[t]
\centering
\begin{tabular}{p{0.15\textwidth}p{0.8\textwidth}}
$category_R$ & Enforces a "type" on clause subset $R$.\\
$exception_R$ & Refines a category.\\
$requirement_R$ & Specifies a precondition for clause subset $R$.\\
$obligation_R$ & Defines a set of events that must occur in the future (in all conflicting timelines) for this action to be compliant in clauses $R$.\\
$permits_R$ & $category_R \wedge \neg exception_R\wedge requirement_R \wedge obligation_R$\\
$forbids_R$ & $category_R \wedge \neg exception_R \wedge (\neg requirement_R \vee \neg obligation_R)$\\
$not\_applicable_R$ & $\neg category_R \vee exception_R$
\end{tabular}
\caption{\label{tab:transmission-rules}Description of the rules for transmission of PHI. All rules are replicated from \cite{lam2009formalization}, except $obligation_R$, which we have added to support transmission constraints over time.}
\end{table}

\section{Querying Generated Narratives}

After encoding the privacy policy, we can begin asking questions about possible consequences of a policy (Box 4 of Figure~\ref{fig:system}). One application of our work is to present unexpected consequences or edge cases to the individual. However, since the open-ended generation of scenarios and their interestingness is subjective, we focus on answering the specific queries enumerated in the motivating example.

\paragraph{\ref{q:alice-phi-third-parties} and \ref{q:alice-phi-third-parties-no-authorization}} Encoding Alice's query is straightforward; we do not list it here. Critically, there exists an action \texttt{information\_sold} in our encoding. The only new information we need to specify in order to answer this query specifically is a fluent asserting that \texttt{information\_sold} should be true. This is done by forbidding the negation of \texttt{information\_sold}. This allows us to filter out narratives in which the condition doesn't hold, without defining the predicate to be true (i.e., the predicate must arise naturally from other parts of the program). For~\ref{q:alice-phi-third-parties-no-authorization}, we simply add one line to Alice's query, which restricts the set of of possible answers returned.

We use output templates to translate the tool output into a human readable format for \ref{q:alice-phi-third-parties}. One possible outcome consists of two events: (1) Alice authorizes the hospital to sell her PHI to an advertiser, then (2) the hospital sells Alice's PHI to that advertiser. 

\paragraph{\ref{q:bob-employer-phi}} Figure~\ref{fig:antagonistic} depicts the tool output to \ref{q:bob-employer-phi}: if Bob quits his job before the info is requested, the doctor cannot disclose the info to the employer. The planner also says that the action of disclosing the information is permitted by \S164.512(b)(v)(A). This graph of possible paths is an an example instance of Box 5 of Figure~\ref{fig:system}.

\section{Related Work}
Formal modeling for policies is not new; while legal text often contains ambiguities and cross-references \cite{bench1987logic}, its structured language lends itself readily to a logical representation. To this end, various modeling approaches have been used, including defeasible logic to handle overlaps and conflicts in legal text \cite{antoniou1999modeling,governatori2009defeasible} and deontic logic to represent permissions and obligations \cite{valente1995line}. For a survey of other related work in this area, we refer the reader to \citet{otto2007addressing}.

For privacy-specific legal modeling, \citet{barth2006privacy,barth2007privacy} present LPU (Logic of Privacy and Utility), a formalization of ``Contextual Integrity,'' applied in the context of business processes to ensure privacy. \citet{deyoung2010experiences} build on that work with PrivacyLFP logic. Their logic supports obligations and temporal reasoning. They use it to produce a comprehensive formalization of transmission-related HIPAA clauses. \citet{lam2009formalization} describe a system for verifying HIPAA compliance of messages sent by hospital employees. They present pLogic, an executable framework implemented as a Datalog program. Their formalization does not support ensuring future obligations. \citet{chowdhury2013privacy} present a policy specification language based on first-order linear temporal logic, along with an algorithm for static policy analysis of policies like HIPAA.

HCI researchers have focused on presenting privacy policies to users in a more comprehensible way. P3P \cite{cranor2003p3p, cranor2006user} was a system for allowing websites to declare the intended usage of collected information. The browser could then block any cookies that would conflict with user privacy settings. \citet{lin2012expectation} use crowdsourcing to capture user's expectations of privacy in the context of mobile apps. They describe a privacy summary interface that emphasizes places where these expectations are broken. \citet{kelley2009nutrition} present a ``nutrition label" for privacy, displaying in a grid what categories of information are accessed to what extent, drawing from the FDA's nutrition facts panel.

These efforts focus on presenting static information to the user in a usable manner. Our system enables a richer exploration of regulations through narratives. Our work fundamentally differs from prior work due to the combination of formalization and narrative generation.

\paragraph{Concluding Remarks}
The goal of this work is not to create another policy modeling logic. It fundamentally differs from the existing efforts, in that it is a \textit{generative} system, as opposed to a \textit{compliance verification} system. While it can be used to verify compliance of a given trace of actions (it does so by attempting to generate a matching narrative), its strength lies in enabling exploration of the encoded regulation through scenarios involving intentional actors.
Our vision with this project is to create tools to enable users to better understand privacy policies, empowering them to make an informed decision on when and with whom to share their data. At the same time, we want user expectations to guide policy design. We believe this work is an important first step in that direction.

\begin{acks}
This material is based upon work supported by the National Science Foundation (CNS \#1846122) and Northeastern University. Any opinions, findings and conclusions or recommendations expressed in this material are those of the author(s) and do not necessarily reflect the views of the National Science Foundation.
\end{acks}

\bibliographystyle{ACM-Reference-Format}
\bibliography{references}

\end{document}